\documentstyle[epsfig]{aipproc}

\begin{document}
\title{Lepton asymmetry effect on neutrino oscillations and
primordial $^4\!He$.}

\author{Daniela Kirilova$^*$ and Mihail Chizhov$^{\dagger}$}
\address{$^*$Institute of Astronomy, Sofia, Bulgaria; e-mail: 
dani@libra.astro.bas.bg\\
$^{\dagger}$Theory Division, CERN, Geneva, Switzerland and
Sofia University, Sofia, Bulgaria}

\maketitle

\begin{abstract}
We analyze the effects
of lepton asymmetry on neutrino oscillations and on
cosmological nucleosynthesis with active-sterile oscillating
 neutrinos.
It is shown that small lepton asymmetries, $L<0.01$, whose
direct kinetic effect on nucleosynthesis is negligible, still
effect nucleosynthesis considerably through their influence
on oscillating neutrinos. Two different cases of lepton asymmetry 
are discussed: an initially present  and a dynamically
generated in oscillations. 
{\it Dynamically generated in resonant oscillations asymmetry} 
at small mixing angles 
suppresses oscillations, hence, the
nucleosynthesis bounds on neutrino mass differences  
at small mixings are relaxed. 
{\it Initially present asymmetry} may  suppress or enhance
oscillations. The enhancement 
is a result of interchanging resonances between
neutrino and antineutrino ensembles due to resonance waves
passing through the neutrino and antineutrino spectrum.
Updated nucleosynthesis bounds on neutrino oscillation
parameters accounting for lepton asymmetry are presented.  
\end{abstract}

\section{\bf Introduction}
\label{s1}

The direct kinetic effect of a considerable lepton asymmetry
$L>0.01$ (either
initially present or dynamically generated) on cosmological
nucleosynthesis (CN) has been 
investigated in
\cite{lepCN}, and cosmological constraints on its value have been
obtained.
The  effect of neutrino oscillations on CN was also studied  
and stringent bounds on oscillation parameters were 
obtained \cite{oscCN,D88,NU96,PR,res,dubna}.
Neutrino oscillations,  proceeding  in the primordial plasma
during CN epoch, can effect $^4\!He$ yield by 
(a) bringing  additional degrees of freedom
into the primordial heat bath\cite{oscCN},
(b) depleting the neutrino 
and antineutrino number densities\footnote{The effect was
estimated for equilibrium~\cite{oscCN} and
nonequilibrium~\cite{D88} oscillations and numerically
calculated for nonequilibrium oscillations in~\cite{NU96}.}
thus slowing the weak interactions 
$\Gamma_w\sim N_{\nu} E^2_{\nu}$,
(c) distorting the neutrino and antineutrino
spectrum~\cite{D88,NU96},
(d) producing  neutrino-antineutrino
asymmetry\footnote{The asymmetry growth was
estimated to be possible for
great mass differences $\delta m^2>10^{-5}$ eV$^2$
\cite{create}, however for $\delta m^2<10^{-7}eV^2$,
a considerable  asymmetry growth was registered in precise
numerical studies~\cite{NU96}.}~\cite{NU96,create}, 
which on its turn
influences the evolution of the neutrino and antineutrino ensembles 
and the oscillation pattern~\cite{NU96,new,res}. 

In this work we discuss the simultaneous effect
of a lepton asymmetry and neutrino oscillations on CN.
Two different cases of lepton asymmetry
are analyzed: an initially present and
generated in oscillations asymmetry.
We have studied the role of a lepton asymmetry on CN with
oscillations, effective after electron neutrino decoupling.  
We have shown by a numerical analyses of the kinetics of 
nucleons and the oscillating neutrinos and antineutrinos 
in CN epoch,  that  much
smaller asymmetries 
$L<<0.01$ exert considerable {\it indirect effects} on CN
through oscillations, due to the fact that even very small
asymmetries change
the medium induced neutrino potential energy and influence
the evolution of the
oscillating neutrinos. 
The prejudice that in order to influence
nucleosynthesis the neutrino asymmetry first must grow to a
considerable value $L>0.01$, is not applicable for the case of
nucleosynthesis {\it with oscillations}.\footnote{Really in the
case {\it without oscillations} the asymmetry has a direct
sign-dependent  effect on the kinetics of
the nucleons, and hence, on the helium
yields. Besides, when large enough it also contributes to the
  Universe energy density thus increasing the expansion rate 
$H(t)$ and changing
primordial helium-4 yield $Y_p$~\cite{lepCN}.}

Lepton asymmetry influences CN with oscillations in several
ways: The neutrino and antineutrino ensembles
evolve differently in comparison with the case without
asymmetry, i.e neutrino number densities, their depletion and
spectrum distortion are changed. Also due to asymmetry term  
neutrino and antineutrino become strongly coupled and evolve 
differently. 
Lepton asymmetry  changes as well  the oscillation
pattern, i.e. leads to an enhancement or suppression of
oscillations. All these resultt into changed light elements yields,  
 compared with the oscillatory case without 
asymmetry.
This indirect  asymmetry influence on
CN is considerable~\cite{NU96,PR,new,NP,now}. However,
inorder to reveal it asymmetry should be  
considered selfconsistently with the neutrinos and nucleons
evolution. The results obtained without the account of the
indirect effects differ by many orders of magnitude from the real 
picture~\cite{new}.\footnote{Works considering
asymmetry effect on CN in case of oscillations
\cite{wrong,FVlast}, providing rough estimates of the asymmetry
growth and its effect on CN, will become more reliable
when a proper account for the indirect asymmetry effects
on CN during the full evolution of the asymmetry is provided in
a selfconsistent analysis of the neutrino and nucleons evolution
for each momentum.} 

Numerical analysis,  
accounting for the asymmetry effect
selfconsistently with
neutrino  and  nucleons 
evolution,
was provided in refs.~\cite{PR,new,res,dubna}. See also 
ref.~\cite{dolasym}, where an precise analytical study of the 
asymmetry evolution, accounting for its back effect on oscillating 
neutrinos, was proposed.  
In the nonresonant case the
oscillations produced asymmetry was shown to have a
negligible role in CN~\cite{PR}. However, in the resonant
oscillation case the
asymmetry effect on CN is
considerable~\cite{NU96,new,res,dolasym,dubna}. 
 
In general, dynamically produced asymmetry  suppresses
oscillations~\cite{NU96,new},
which leads to less overproduction of helium-4 in comparison
with  CN with oscillations but without an
 asymmetry account. Hence, the bounds on oscillation
parameters are alleviated at small mixing
angles~\cite{new,res,dubna}. We present in sec.~\ref{s3} the 
updated cosmological constraints for electron-sterile 
oscillation case,
accounting precisely for the 
oscillations generated asymmetry.

The effect of small initial lepton
asymmetries ($10^{-10}<L<10^{-4}$) on
 CN with nonresonant  active-sterile oscillations 
was precisely studied in 
~\cite{PR,NP,now}. It was found that asymmetry is able to enhance
oscillations, besides
its well known ability to suppress them~\cite{NP}, 
 thus  leading correspondingly  to an over-  or
under-production of helium. 
Our analysis has shown
(see sec.~\ref{s4}.) that the initially present asymmetry 
is able to alleviate CN bounds at large mixings and 
to tighten the bounds at small mixings  (see also
refs.~\cite{NP,now}).

In the next section we present the precise kinetic approach 
for the description of lepton asymmetry effects on CN in the 
presence of oscillations.

\section{The kinetics}
\label{s2}

We have  used for the precise analysis
the synthesis of helium-4.
According to the standard CN  the
primordial helium yield depends on two compelling   
processes, determinning the
nucleons freezing:  Universe's cooling, $H(t)\sim
g_{eff}T^2$ and
 weak processes, $\Gamma_w$. Three neutrino flavours, 
zero lepton asymmetry and
equilibrium neutrino number densities and spectrum are assumed.
In the case of CN with oscillations and with lepton asymmetry, 
all these assumptions do not work. According to (a)-(d): neutrino 
oscillations change 
the number of neutrino flavours; they may produce nonequilibrium 
neutrino number densities (particularly the electron neutrino 
density may be considerably reduced in favour of the sterile neutrino 
density); the neutrino spectrum may be distorted in  
active-sterile oscillations; besides, even if initially the lepton 
asymmetry is assumed zero resonant oscillations may lead to  
a considerable growth of the asymmetry, which on its turn effects 
neutrinos and nucleons evolution via oscillations.    
This nonequilibrium picture is hard to describe analytically.
\footnote{There exist in literature numerous analytical 
studies, like refs.~\cite{wrong,FVlast} 
proposing different
schemes for applying neutrino oscillations to solve different
astrophysical or else problems in which schemes a central role
is played by 
a lepton asymmetry (initially present or dynamically generated
in oscillations)  suppressing oscillations. We would like to
warn that  these publications provided too
 rough estimations both of the asymmetry evolution and of its
effect on primordial nucleosynthesis.
The asymmetry evolution  was usually
semianalitically
described using different sorts of simplifying assumptions, 
in general not applicable in the nonequilibrium situation of 
active-sterile oscillations. The
neutrino and antineutrino evolution was described in
terms of particle densities (not particle density
matrix) and
almost equilibrium spectrum (which is not an 
adequate
description of oscillationary phenomena when the growth of
asymmetry is considerable). The asymmetry effect on  nucleosynthesis 
was reduced only to $L$ kinetic effect on CN. Therefore, its role
in nucleosynthesis was accounted for
after it has grown ``enough''and hence, the indirect effect of
asymmetry during its growth has been neglected. 
The asymmetry back effect on oscillations, in case considered at
all, was assumed to be only towards suppressing oscillations.}

Inorder to account for the different effects of oscillations and
asymmetry on CN a  selfconsistent 
 numerical analysis of
the kinetics of the oscillating neutrinos, the  nucleons freeze-out
and the asymmetry evolution is necessary. In our analysis the 
set of
the following coupled integro-differential equations describing the
evolution of the neutrino $\rho$ and neutron number densities  $n_n$
was solved simultaneously and selfconsistently:

\begin{eqnarray}
{\partial \rho(t) \over \partial t} &=&
H p_\nu~ {\partial \rho(t) \over \partial p_\nu} +
\nonumber\\
&+& i \left[ {\cal H}_o, \rho(t) \right]
+i \sqrt{2} G_F \left({\cal L} - Q/M_W^2 \right)N_\gamma
\left[ \alpha, \rho(t) \right]
+ {\rm O}\left(G_F^2 \right),
\nonumber\\
\nonumber\\
{\partial\bar\rho(t) \over \partial t} &=&
H p_\nu~ {\partial \bar\rho(t) \over \partial p_\nu} +
\nonumber\\
&+& i \left[ {\cal H}_o,\bar\rho(t) \right]
+i \sqrt{2} G_F \left(-{\cal L} - Q/M_W^2 \right)N_\gamma
\left[ \alpha, \bar\rho(t) \right]
+ {\rm O}\left(G_F^2 \right),
\label{kin}
\end{eqnarray}

\begin{eqnarray}
&&\left(\partial n_n / \partial t \right)
 = H p_n~ \left(\partial n_n / \partial p_n \right) +
\nonumber\\
&& + \int {\rm d}\Omega(e^-,p,\nu) |{\cal A}(e^- p\to\nu n)|^2
\left[n_{e^-} n_p (1-\rho_{LL}) - n_n \rho_{LL} (1-n_{e^-})\right]
\nonumber\\
&& - \int {\rm d}\Omega(e^+,p,\tilde{\nu}) |{\cal A}(e^+n\to
p\tilde{\nu})|^2
\left[n_{e^+} n_n (1-\bar{\rho}_{LL}) - n_p \bar{\rho}_{LL}
(1-n_{e^+})\right].
\end{eqnarray}

\noindent where $\alpha_{ij}=U^*_{ie} U_{je}$, mixing just in the electron
sector was assumed $\nu_i=U_{il}\nu_l (l=e,s)$.
$p_\nu$ is the momentum of electron neutrino,
 $n$ stands for the number density of the interacting particles,
${\rm d}\Omega(i,j,k)$ is a phase space factor and  ${\cal A}$ is the
amplitude of the corresponding process.

These equations  provide simultaneous account of the different
competing processes,
namely: neutrino oscillations, Hubble expansion and weak interaction
processes.
${\cal H}_o$ is the free neutrino Hamiltonian.
The `nonlocal' term $Q$ arises as an $W/Z$ propagator effect,
$Q \sim E_\nu~T$.
${\cal L}$ is proportional to the fermion asymmetry of the plasma
and is essentially expressed through the neutrino asymmetries
${\cal L} \sim 2L_{\nu_e}+L_{\nu_\mu}+L_{\nu_\tau}$,
where
$L_{\mu,\tau} \sim (N_{\mu,\tau}-N_{\bar{\mu},\bar{\tau}})/ N_\gamma$
and $L_{\nu_e} \sim \int {\rm d}^3p (\rho_{LL}-\bar{\rho}_{LL})/N_\gamma$.

The neutron and proton number  
densities, used in the kinetic equations for neutrinos,
are substituted
from the numerical calculations of eq.~(2). On the other hand,
$\rho_{LL}$ and $\bar{\rho}_{LL}$
at each integration step of eq.~(2) are taken  from the
simultaneously  performed integration of the set of equations
(1).

The equations are for the neutrino and neutron number densities
in
momentum space. This allows to account precisely for the
spectrum
distortion effect and neutrino depletion effects of
oscillations,
as well
as to follow the evolution of the neutrino asymmetry and its back 
effect at each neutrino 
momentum.

In our  numerical analysis the spectrum
distortion was described by 1000 bins for 
the
nonresonant case and by  5000 bins for the resonant case. 
In case the spectrum was described by $N$ bins,
a system of $6N+1$ coupled integro-differential equations following 
from (1) and (2), was numerically solved.

The numerical analysis was provided
for the characteristic temperature interval 
$\left[2\mbox{ MeV},0.3\mbox{ MeV}\right]$ and the 
full set
of oscillation parameters of the active-sterile oscillation model 
\cite{NU96}. 
We calculated precisely the $n/p$-freezing, 
essential  for the
production of helium, till temperature $0.3$ MeV, and accounted 
adiabatically
for the following decays of neutrons till the start of 
nuclear reactions
at about $0.1$ MeV.

\section{\bf Dynamically generated lepton asymmetry}
\label{s3}

Our numerical analysis showed that in the resonant oscillation 
case the dynamical neutrino-antineutrino asymmetry grows    
up to 4 orders of magnitude.
I.e.\ starting with asymmetries of the order of the
 baryon one it reaches
maximum a value $10^{-5}$. Hence, having in mind this small value   
the registered asymmetry effect is totally due
to its {\it indirect} influence on CN via oscillations.
Dynamically produced asymmetry at small mixing angles  
suppresses
oscillations,
which leads to less overproduction of $^4\!He$ in comparison
with  CN with oscillations but without an
 asymmetry account. Hence, the cosmological constraints 
 on oscillation
parameters are alleviated at small mixing
angles.                        

\tabcolsep=1.5mm
\begin{figure}[t]
\epsfig{file=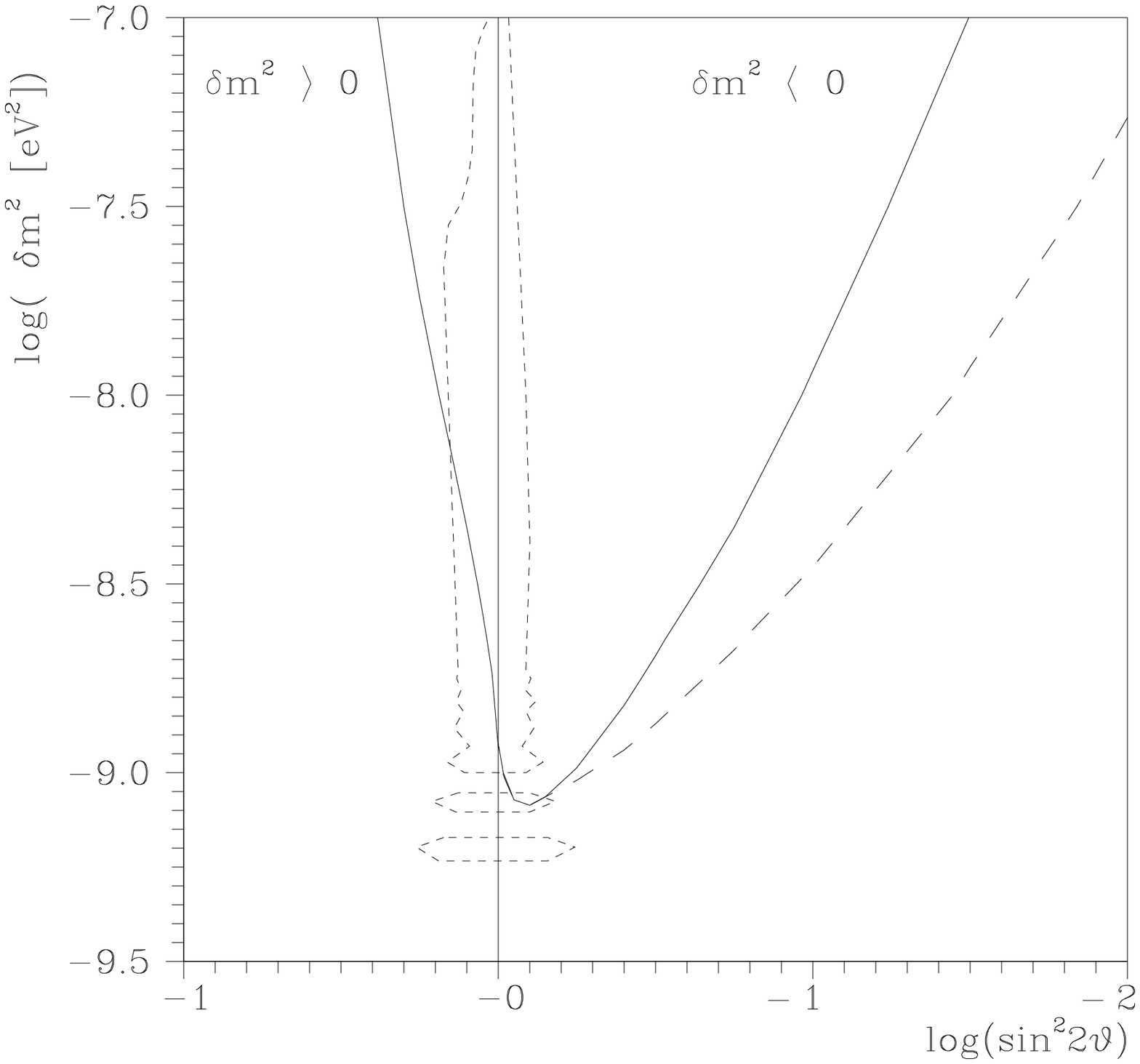,height=5cm,width=4.9cm}
\epsfig{file=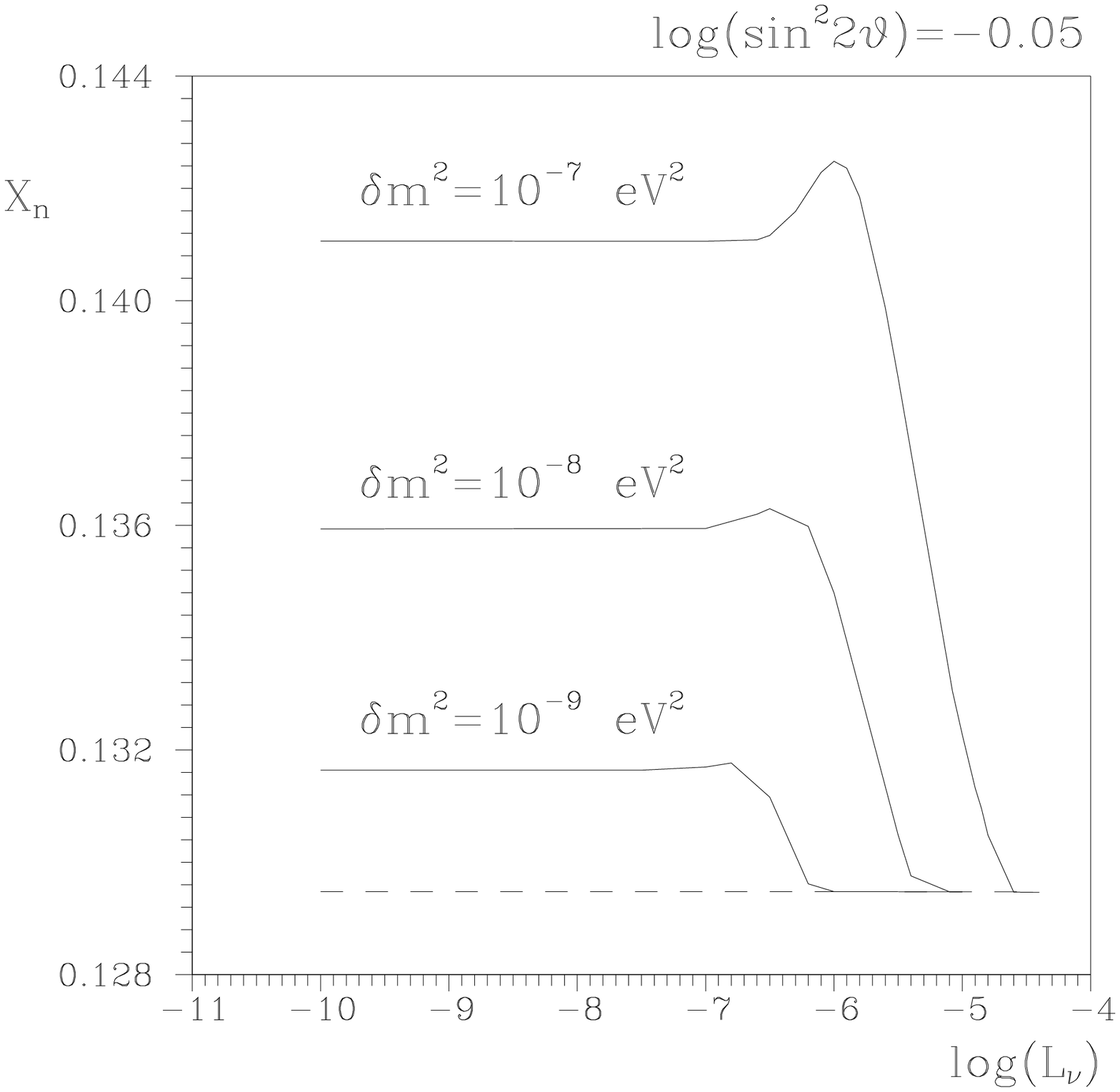,height=5cm,width=4.8cm}
\epsfig{file=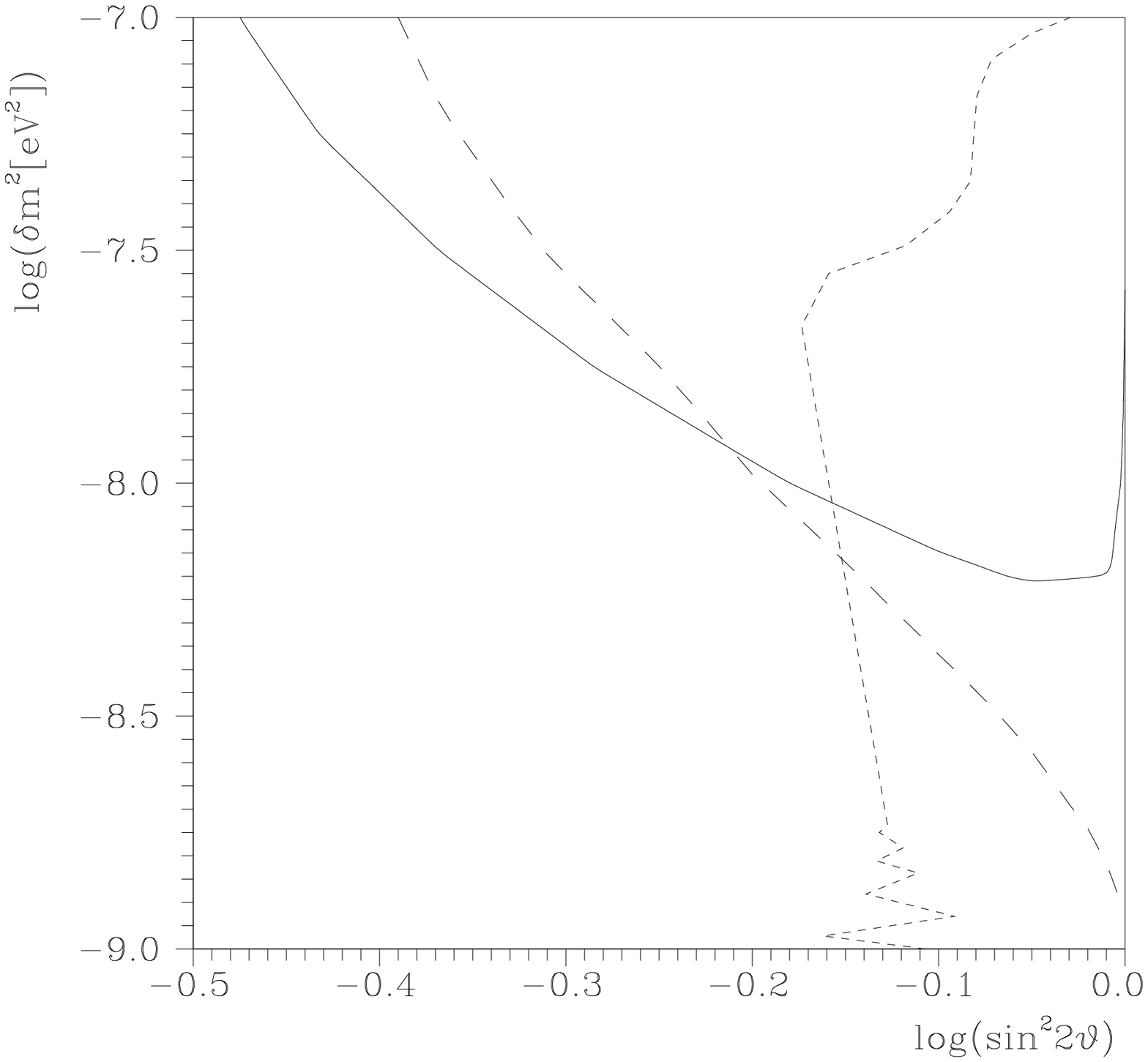,height=5cm,width=4.9cm} 
\mbox{\hspace{-0.6cm}}
\vspace{0.1cm}
\begin{tabular}{p{4.8cm}p{4.8cm}p{4.8cm}}
\baselineskip1pt
{\scriptsize
{\bf Figure 1.} 
Cosmological constraints for the electron-sterile neutrino
oscillations, are presented by the solid curves 
$Y_p=0.24$.
The dashed curve shows the contour
without asymmetry account. 
The dotted curve shows solar neutrino LOW solution. 
\mbox{\hspace{10cm}}
} &
\baselineskip1pt
{\scriptsize
{\bf Figure 2.} 
The dependence of the neutron number density relative to
nucleons
$X_n=N_n/(N_p+N_n)$ for the case of oscillations  with
$\sin^2(2\vartheta)=10^{-0.05}$ 
on the value of the  initial asymmetry is plotted \cite{NP}.
\mbox{\hspace{10cm}}
} &
\baselineskip1pt
{\scriptsize   
{\bf Figure 3.}
On the $\delta m^2-\vartheta$ plane the
isohelium contours $Y_p=0.24$  calculated in the discussed
model of CN with neutrino oscillations
and initial lepton asymmetries  $L=10^{-6}$ (solid curve)
and $L=10^{-10}$ are shown. The dotted curve denotes LOW 
solution. 
\mbox{\hspace{10cm}}
}
\end{tabular}
\vspace{-1.1cm}
\end{figure}

The updated constraints on active-sterile
neutrino oscillations, precisely accounting for the
asymmetry generation, spectrum distortion
and the depletion of the neutrinos, are presented in fig.~1.
The
plots correspond to $Y_p=0.24$.
The net {\it indirect} asymmetry effect on CN is given 
for the resonant case 
by the 
difference between the dashed curve (without asymmetry account) 
and the solid one.
Due to asymmetry growth account, 
and the corresponding suppression of oscillations, 
$Y_p$ overproduction is
not so strongly expressed at small mixing angles, hence 
CN constraints are
alleviated  at small mixing angles for $\delta m^2 <0$.

In the resonant case the cosmological constraints at large mixings
are 
$|\delta m^2| \le 8.2\times 10^{-10}$ eV$^2$.
In the nonresonant case $\delta m^2 >0$
an analytical fit to the exact constraints is:
$\delta m^2 (\sin^2 2\vartheta)^4\le 1.5\times 10^{-9}$ eV$^2$.
 The constraints in both cases are 
 strengthened compared to  the previous ones~\cite{oscCN}
due to the precise account of the spectrum distortion
and to the exact kinetic approach
to the neutrinos and nucleons evolution.

According to these constraints the LOW active-sterile solution
to the solar neutrino problem, which is favoured by the
analysis of the recent experimental data of total measured rates
and day and night spectrum measured by
SuperKamiokande~\cite{suzuki} is almost completely
 excluded~\cite{now}.

\section{\bf Initially present lepton asymmetry}
\label{s4}

The role of initially present relic  asymmetries on  CN with
nonresonant  active-sterile oscillations was precisely 
studied, following the 
lines of work described in sec.~\ref{s2}. 
A wide range of  $L$ values $[10^{-10}-10^{-4}]$ was
analyzed. Such small asymmetries have only indirect  effects on
CN. 
On fig.2 the dependence of the
produced helium, in a CN model with oscillations  
on the initial asymmetry is plotted. 
It was found that the asymmetry is able also to
{\it enhance} oscillations, besides
its well known ability to suppress them. 
 The enhancement is a
synthetic  effect of a resonant wave passing through neutrino
spectrum till lepton asymmetry changes sign and followed by a
similar `spectrum' resonance at the antineutrino ensemble and
vice versa~\cite{NP}.
Hence,
this
enhancement has a complex spectral character and could be
revealed  only by a 
 precise kinetic approach to the oscillation
problem. The influence of the asymmetry should not be reduced 
only to oscillations suppression, as usually believed.  
Depending on the concrete values of
oscillations parameters the  asymmetry  may suppress, enhance
or not influence
oscillations, thus  leading correspondingly  to an under- ,
over-production of helium   
or not change its abundance.
 Hence,
in order to judge the real asymmetry effect on CN,
precise numerical analysis for the concrete oscillation
parameters and $L$ values is obligatory. Initially present
asymmetry may relax CN bounds at large mixings and
tighten the bounds at small mixings. 

Qualitatively, for
oscillations affective after the freeze-out of the electron
neutrino, the asymmetry effect is as follows:
$L<10^{-7}$ has negligible effect on CN; 
 $10^{-7}<L<10^{-5}$ enhances
oscillations due to the
spectrum wave  resonance~\cite{NP}, 
resulting  into an
enhanced overproduction of helium-4;
$L>10^{-5}$ leads to a
suppression of oscillations and  relaxation of the
CN bounds (fig.~2).

In fig.~3 the isohelium contours $Y_p=0.24$ are  presented, 
for different $L$: $L=10^{-6}$ and $L=10^{-10}$. For small
mixing angles  the asymmetry $L=10^{-6}$ enhances oscillations,
which reflects into stronger  bounds on oscillation
parameters, while for large mixings  
this asymmetry suppresses 
oscillations and  CN
bounds are weakened compared with $L=10^{-10}$ case.

A similar investigation, of initial asymmetry effect on CN 
with oscillations, for the resonant case will be a more 
complicated
task  due to technical problems:  The usual explicit numerical
approach is not applicable
for the description of the asymmetry evolution, because  the
neutrino evolution
equations at resonance have  high stiffness; besides the
resonance case  deserve much greater
number of
bins for the spectrum distortion description. 
To solve the stiff equations numerically, 
implicit methods should be used. For 5000 bins of the spectrum 
a system of 30000 equations
describing the neutrino  evolution should be 
solved simultaneously.   
However, this investigation is interesting, 
as far as such small values of
the initial relic asymmetry  are not excluded 
neither from observations
nor
from some profound theoretical principle.

In conclusion we would like to stress that 
small asymmetries, initially present or dynamically generated,
influence CN thanks to their backfeed effect on oscillating 
neutrinos.The account of spectrum distortion of the oscillating neutrinos
as well as the  selfconsistent account of 
neutrinos and  nucleons evolution is essential for revealing the
indirect effect of small lepton asymmetries. 
Therefore, a precise kinetic approach
should be provided when analyzing the effect of 
lepton asymmetry on CN with oscillations.\\

We are glad to thank the organizing committee of CAPP2000,
for the stimulating atmosphere of the conference and
for the financial support of the participation of D.K.
M.C. thanks Theory Division of CERN where this work was prepared.


\begin{references}

\bibitem {lepCN} Wagoner R. V., Fowler W. A., and
Hoyle F., {\it Astrophys. J.} {\bf 148}, 3 (1967);
Terasawa N., and Sato K., {\it Prog. Theor.
Phys.} {\bf 80}, 468 (1988) and the references there in.

\bibitem{oscCN} Dolgov A. D., {\it Sov. J. Nucl. Phys.}
{\bf 33}, 700 (1981); Barbieri R., and Dolgov A.,
{\it Phys. Lett.} {\bf B 237}, 440 (1990);
{\it Nucl. Phys.} {\bf B 349}, 743 (1991);
Enqvist K., Kainulainen K., and Thomson M.,
{\it Nucl. Phys.} {\bf B 373}, 498 (1992);
Dolgov A. D., hep-ph/0006103.

\bibitem{D88} Kirilova D. P., {\it JINR preprint} E2-88-301,
1988.

\bibitem{NU96} Kirilova D. P., and Chizhov M. V.,
in Proc. NEUTRINO 96 Conference, Helsinki, 1996, p. 478;
{\it Phys. Lett.} {\bf B 393}, 375 (1997)

\bibitem{PR} Kirilova D. P., and Chizhov M. V.,
{\it Phys. Rev.} {\bf D 58}, 073004 (1998).

\bibitem{res} Kirilova D. P., and Chizhov M. V.,
{\it Nucl. Phys.} {\bf B 591}, 457 (2000).

\bibitem{dubna} Kirilova D., talk at
Astrophysics Workshop "Hot Points in Astrophysics",
 August 2000, Dubna, Russia, to be published in the proceedings.

\bibitem{create} Foot R., Thomson M. J., and Volkas R. R.,
{\it Phys. Rev.} {\bf D 53}, R5349 (1996);
Shi X., {\it Phys. Rev.} {\bf D 54}, 2753 (1996).

\bibitem{new} Kirilova D. P., and Chizhov M. V., hep-ph/9908525.

\bibitem{NP} Kirilova D. P., and Chizhov M. V.,
{\it Nucl. Phys.} {\bf B 534}, 447 (1998).

\bibitem{now} Kirilova D., talk at NOW2000 Workshop, 11-16
Sept. 2000, Otranto, Italy, to be published in Nucl. Phys.B.

\bibitem{wrong} Abazajian K., Shi X., and Fuller G. M., astro-ph/9904052;
Shi X., Fuller G. M., and Abazajian K., astro-ph/9908081;
{\it Phys. Rev.} {\bf D 60}, 063002 (1999);
Shi X., and Fuller G. M.,
{\it Phys. Rev. Lett.} {\bf 83}, 3120 (1999).

\bibitem{FVlast}  Foot R., and Volkas R. R.,
{\it Phys. Rev.} {\bf D 55}, 5147 (1997);
{\it ibid.} {\bf D 56}, 6653 (1997); Erratum
{\it ibid.} {\bf D 59}, 029901 (1999);
{\it Phys. Rev.} {\bf D 61}, 043507 (2000);
astro-ph/9811067;
Bell N., Foot R., and Volkas R. R.,
{\it Phys. Rev.} {\bf D 58}, 105010 (1998);
Foot R., {\it Astropart. Phys.} {\bf 10}, 253 (1999);
{\it Phys. Rev.} {\bf D 61}, 023516 (2000);
Di Bari P., Lipari P., and Lusignoli M.,
{\it Int. J. Mod. Phys.} {\bf A 15}, 2289 (2000).

\bibitem{dolasym}Dolgov A. et al. hep-ph/9910444, TAC-1999-018.

\bibitem{suzuki} Suzuki Y., talk at Neutrino 2000, June 2000, Sudbury,
Canada; Gonzalez-Garcia M. C., Pe\~na-Garay C., hep-ph/0009041
talk at Neutrino 2000, June 2000, Sudbury, Canada;  Bahcall J. N.,
Krastev P. I., and Smirnov A. Yu., talk at NOW2000, Sept. 2000,
Otranto, Italy.
\end{references}
\end{document}